\documentclass[pra,twocolumn,showpacs,superscriptaddress,preprintnumbers,amsmath,amssymb,floatfix]{revtex4}
\usepackage{bm}
\usepackage{graphicx}
\usepackage{amsbsy}
\usepackage{amsmath}
\usepackage{amsfonts}
\usepackage{amsthm}
\usepackage{color}
\usepackage{mathrsfs}
\usepackage{graphicx}
\usepackage{dcolumn}
\usepackage{bm}
\usepackage[OT2,T1]{fontenc}
\DeclareSymbolFont{cyrletters}{OT2}{wncyr}{m}{n}
\DeclareMathSymbol{\Sha}{\mathalpha}{cyrletters}{"58}

\newcommand{\bwt}{\begin{widetext}}
\newcommand{\ewt}{\end{widetext}}
\newcommand{\bea}{\begin{eqnarray}}
\newcommand{\eea}{\end{eqnarray}}

\begin{document}

\title{Rephasing processes and quantum memory for light: \\reversibility issues and how to fix them}

\author{Sergey A. Moiseev}
\email{samoi@yandex.ru}

\affiliation{Kazan Physical-Technical Institute of the Russian Academy of Sciences,10/7 Sibirsky Trakt, Kazan, 420029,
Russia}
\affiliation{Institute for Informatics of Tatarstan Academy of Sciences, 20 Mushtary, Kazan, 420012, Russia}

\author{J.-L. Le Gou\"{e}t}
\email{jean-louis.legouet@lac.u-psud.fr}

\affiliation{Laboratoire Aim\'{e} Cotton, CNRS-UPR 3321, Univ. Paris-Sud, B\^{a}t. 505, 91405 Orsay cedex, France}

\date{\today}

\begin{abstract}

Time reversibility is absent from some recently proposed quantum memory protocols such as Absorption Frequency Comb (AFC). Focusing on AFC memory, we show that quantum efficiency and fidelity are reduced dramatically, as a consequence of non-reversibility, when the spectral width of the incoming signal approaches the memory bandwidth. Non-reversibility is revealed through spectral dispersion, giving rise to phase mismatching. We propose a modified AFC scheme that restores reversibility. This way, signals can be retrieved with excellent efficiency over the entire memory bandwidth. This study could be extended to other quantum memory rephasing schemes in inhomogeneously broadened absorbing media.

\end{abstract}

\maketitle


\section{Introduction}
\label{sec:introduction}

Rephasing processes in inhomogeneously broadened absorption lines~\cite{hahn1950,kopvil1963,kurnit1964} offer an attractive way to store and retrieve a large number of optically carried temporal modes~\cite{fern1955,elyu1979,zuikov1980,mossberg1982}. This property raises increasing interest in the quest for multi-mode quantum memories (QMs) for light~\cite{lvovsky2009, tittel2010, simon2010}.

In the first proposed rephasing QM scheme, known as Controlled Reversible Inhomogneous Broadening (CRIB)~\cite{moiseev2001,moiseev2003,nilsson2005,kraus2006}, the retrieval step is devised to be the exact time reversed copy of the storage phase. Perfect reversibility is preserved even with broadband signals spanning the entire absorption line, provided all the incoming spectral components are completely captured in the memory material. The required inhomogeneous broadening is generated by an external electric field through Stark effect. The frequency detuning is reversed by Stark switching. Hence atomic coherences can be brought back in phase together and be able to restore the original signal. To satisfy the time and space symmetry requirement, the signal must be recovered in backward direction, where efficiency can reach 100\%. To this end, the initially excited coherences have to be converted into spin or hyperfine coherences and back. In forward direction, retrieval efficiency drops to less than 54\%.

The practical implementation of CRIB led to the emergence of a novel scheme, known as Gradient Echo Memory (GEM)~\cite{alexander2006,hetet2008,hetet2008b}, where the absorption line is \textit{not} inhomogeneously broadened \textit{locally}. Instead inhomogeneous broadening only appears through integration over the material depth. With GEM, 100\% efficiency is expected in forward direction, in spite of imperfect reversibility. Actually, record quantum efficiency has been reported recently in solid~\cite{hedges2010} and gaseous~\cite{hosseini2011} media. The lack of reversibility is revealed by fidelity reduction~\cite{hetet2008,moiseev2008}. Experimentally, with retrieval in forward direction, one no longer needs ancillary spin or hyperfine coherences, as in CRIB. Hence GEM works with two-level atoms.

Alternative rephasing techniques have been proposed recently~\cite{deried2008,afzelius2009,moiseev2011,mcauslan2011,damon2011}. They avoid the frequency detuning reversal step. Among those new protocols, Absorption Frequency Combs (AFC) have already given rise to a large harvest of experimental promising results. In AFC-based memories, the signal is captured by an evenly spaced spectral array of absorption pikes. As a result, coherences are automatically phased back together after a delay given by the inverse spacing of the absorbing pikes. In contrast with CRIB, the revival step is not time reversed with respect to the storage step. Actually, the signal is restored with the same time order as the original one. Using AFC, much wider bandwidth with much larger temporal multimode capacity has been demonstrated than with CRIB and GEM~\cite{usmani2010,bonarota2011}. In addition, for the first time in a solid, storage and retrieval of photonic entanglement has been demonstrated using AFC~\cite{clausen2011,sagla2011}. However those results have been obtained with rather modest quantum efficiency, presumably because of practical difficulties in the absorption pike array preparation.

Beyond practical limitations, one should question the lack of temporal symmetry or reversibility in the latter processes. In the present paper we examine this question in the specific case of AFC-based memories. In Section~\ref{sec:capturing}, we consider the capture of an incoming quantum field by a finite-bandwidth AFC. In section~\ref{sec:recovery}, we express the recovered signal and we analyze the negative impact of dispersion effects on AFC quantum efficiency and fidelity. We also delineate the difference between CRIB and AFC regarding reversibility. In Section~\ref{sec:MAFC}, in the light of this analysis, we propose a modified AFC scheme (MAFC) providing almost 100{\%} quantum efficiency and perfect fidelity over the whole spectral range of the AFC filter. Finally we briefly consider the extension of this method to other non-reversible QM for light protocols.

\section{Capturing the incoming signal}\label{sec:capturing}

Following the basic scheme of photon echo QM, we consider an ensemble of three-level atoms initially prepared in the ground state
$\left|1\right\rangle =\prod\limits_{j=1}^N {\left| 1 \right\rangle _j } $. The quantum dynamics of the light-matter system is described by linearized
Maxwell-Bloch equations. In the weak field limit, the $j$th atom coherence $\mathbf {S}(\tau,\Delta _j ,z_j )=\left| 1 \right\rangle _{jj} \left\langle
2 \right|$ is driven by the signal field $\mathbf{A}_p (\tau ,z)$ according to:
\begin{align}
\label{eq1}
\partial _\tau \hat {S}(\tau ,\Delta _j ,z)= & -i(\omega _o +\Delta _j )\mathbf{S}(\tau ,\Delta _j ,z)
\nonumber  \\&
+ig\mathbf {A}_p (\tau ,z)\exp \{-i\omega _p \tau \},
\end{align}
where $\omega _o $ and $\omega _p $ respectively denote the atomic transition central frequency and the signal field carrier frequency, and $\tau
=t-z/c$ represents time in the moving frame. The radiative reaction of the atoms to electromagnetic excitation is expressed at a macroscopic level,
involving average over many atoms and substituting a continuous medium to the discrete atom distribution. This scale change is mediated by the
macroscopic coherence:
\begin{equation}
\mathbf{S}(\tau ,z)=\int {d\Delta C(\Delta )} \mathbf {S}(\tau ,\Delta ,z),
\end{equation}
where average at position $z$ is performed over the transition frequency distribution $C(\Delta )$. The light field propagation is then described by:
\begin{equation}
\label{eq2}
\partial _z \mathbf {A}_p (\tau ,z)=i\beta \exp \{i\omega _p \tau \}\mathbf
{S}(\tau ,z),
\end{equation}
where $\beta =\pi (n_o S_o )(g^{\ast} /c)$ is function of the atomic concentration $n_o$, the light-beam cross-section S$_{o}$ and the photon-atom
coupling constant $g$.

\begin{figure}
\includegraphics[width=0.4\textwidth]{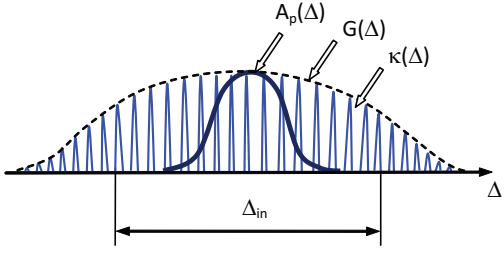}
\caption{AFC filter $C(\Delta )$ uniformly engraved over the entire width of an inhomogeneously broadened absorption line $G(\Delta )$. The initial
inhomogeneous width, the absorbing pike profile and the incoming signal spectrum are respectively denoted as $\Delta _{in}$, $\kappa(\Delta )$and $A_p (\Delta )$.}
\label{Figure1}
\end{figure}

Let us consider an ideal AFC structure, uniformly engraved over the entire available inhomogeneous width, as depicted in Fig.\ref{Figure1}:
\begin{equation}
\label{eq3}
C(\Delta )=\frac{1}{\kappa (0)\delta }\left[ {\Sha\left( {\frac{\Delta }{\delta
}} \right)\otimes \kappa(\Delta )} \right]G(\Delta )
\end{equation}
where $\Sha(x)=\sum\limits_n {\delta (x-n)} $ represents a Dirac comb, and $\kappa(\Delta)$ stands for the profile of each comb tooth. The widths of
$G(\Delta )$ and $\kappa (\Delta )$ are denoted $\Delta _{in} $ and $b$ respectively. They satisfy the condition $b<<\delta <<\Delta _{in} $. Both
$G(\Delta )$ and $\kappa (\Delta )$ are normalized according to~:
\begin{equation}
\label{eq4}
\int {G(\Delta )d\Delta =} \int {\kappa (\Delta )d\Delta =} 1.
\end{equation}
The normalization factor $\left[ {\kappa (0 )\delta} \right]^{-1}\approx b/\delta $ makes $C(\Delta )$ coincide with the initial distribution
$G(\Delta )$ at the top of the comb teeth. Solving the Bloch equation (\ref{eq1}) and summing over $\Delta $ one obtains:
\begin{equation}
\label{eq5}
\mathbf {S}(\tau ,z)=ige^{-i\omega _p \tau }\int_{-\infty }^\tau {d\tau '} \mathbf
{A}_p (\tau ',z)\int {d\Delta C(\Delta )} e^{-i\Delta (\tau -\tau ')}.
\end{equation}
The Dirac comb satisfies the Fourier transform property:
\begin{equation}
\label{eq6}
\int {dx}\Sha\left( x \right)e^{-ixu}=\Sha\left( {\frac{u}{2\pi }} \right).
\end{equation}
Therefore~:
\begin{equation}
\label{eq7}
\int {d\Delta C(\Delta )e^{-i\Delta t}=} \frac{1}{\kappa(0)\delta}\sum\limits_n {\tilde {\kappa }\left( {\frac{2\pi n}{\delta }} \right)}
\tilde {G}\left( {t-\frac{2\pi n}{\delta }} \right),
\end{equation}
where $\tilde {f}\left( t \right)=\int {d\Delta f(\Delta )e^{-i\Delta t}}$. Substitution into Eq.(\ref{eq5}) leads to:
\begin{equation}
\label{eq8}
\mathbf {S}(\tau ,z)=\sum\limits_n {\mathbf {S}^{(n)}(\tau ,z)} ,
\end{equation}
where:
\begin{align}
\label{eq9a}
\mathbf {S}^{(n)}(\tau ,z)= & ige^{-i\omega _p \tau }\frac{1}{\kappa(0)\delta }\tilde{\kappa}\left({\frac{2\pi n}{\delta }}\right)\nonumber\\&
\int_0^\infty {d\tau '} \mathbf {A}_p (\tau -\tau ',z)\tilde {G}\left( {\tau '-\frac{2\pi n}{\delta }} \right).
\end{align}
Because of causality, expressed by the lower boundary in the integral, the sum runs over positive n values only. According to this expression, the
macroscopic coherence revives periodically, with $2\pi/\delta$ time intervals.

If the incoming signal is much shorter than $2\pi /\delta$, the $n$-indexed terms do not overlap. Each one represents the coherence revival at time
$2n\pi/\delta$. The $n=0$ term is involved in the instantaneous signal propagation. Then Eq.(\ref{eq2}) reduces to:
\begin{equation}
\label{eq10}
\partial_z\mathbf{A}_p(\tau ,z)=-\frac{\beta g}{\kappa(0)\delta}\int_0^\infty{d\tau '}\mathbf{A}_p(\tau -\tau ',z)\tilde {G}\left({\tau '} \right).
\end{equation}
where we have replaced $\tilde {\kappa }(0)$ par 1, in accordance with the normalization condition, and we have dropped the $e^{-i(\omega_o-\omega_p
-i\gamma )\tau '}$ factor, very close to unity. The spectral amplitude, defined as $\tilde{\mathbf {A}}_p(\omega ,z)=\int_{-\infty}^\infty {d\tau
}\mathbf {A}_p (\tau ,z)\exp \{i\omega \tau \}$ satisfies the familiar equation:
\begin{equation}
\label{eq11}
\partial _z \tilde {\mathbf {A}}_p (\omega ,z)+i\frac{\omega_p }{2c}\chi (\omega)\tilde {\mathbf {A}}_p (\omega ,z)=0,
\end{equation}
where the real and imaginary parts of susceptibility $\chi (\omega )$ respectively read as:
\begin{equation}
\label{eq12}
\chi '(\omega )=\frac{c}{\omega_p}\alpha(0)\frac{1}{\pi}\mathrm{P}\int_{-\infty }^\infty {\frac{G(\Delta )/G(0)}{\omega-\Delta }}d\Delta ,
\end{equation}
\begin{equation}
\label{eq13}
\chi "(\omega )=-\frac{c}{\omega_p}\alpha(\omega ),
\end{equation}
where $\mathrm{P}$ denotes the Cauchy principal value and the linear absorption coefficient is given by:
\begin{equation}
\label{eq:alpha}
\alpha(\omega)=\frac{2\pi\beta g}{\kappa (0)\delta }G(\omega ).
\end{equation}
One easily solves Eq.(\ref{eq11}) as:
\[\tilde {\mathbf {A}}_p (\omega ,z)=\exp \{-i\frac{\omega_p }{2c}\chi(\omega )z\}\tilde {\mathbf {A}}_p (\omega ,0)\].

With respect to the initial atomic distribution, the absorption coefficient is modified by the factor $[\kappa(0)\delta ]^{-1}\approx b/\delta $. When $G(\Delta )$ is symmetric, $\chi '(0)=0$ and $\vert \chi '(\omega )\vert $ increases almost linearly for $\vert \omega\vert /\Delta _{in} \le 1$.

Contemplating Eqs.(\ref{eq11}-\ref{eq13}), one must remember they rely on the assumption that the incoming signal spectrum is much broader than the
absorbing comb structure. Hence $\chi(\omega)$, representing a coarse graining approximation of the comb susceptibility, is expressed in terms of AFC
envelope $G(\Delta)$. Then $\chi(\omega)$ exhibits the features of an absorption line, with abnormal dispersion within the absorption profile. Should
the absorption coefficient $\omega_p\chi"(\omega)/c$ be large enough, this may result in negative group velocity and so-called superluminal propagation~\cite{wang2000}.

As an illustration, let both $G(\Delta )$ and $\kappa(\Delta )$ be given a gaussian profile according to:
\[G(\Delta )=\sqrt{\frac{\zeta}{\pi}}\frac{1}{\Delta_{in}} \exp \left({-\zeta\frac{\Delta ^2}{\Delta _{in}^2}}\right)\]
\[\kappa(\Delta )=\sqrt{\frac{\zeta}{\pi}}\frac{1}{b} \exp \left({-\zeta\frac{\Delta ^2}{b^2}}\right)\]
where $\zeta=4\mathrm{Ln}(2)$, and $\Delta _{in}$ and $b$ stand for the full width at half maximum (FWHM) of the respective shapes. Assuming $\delta >>b$, substitution in Eqs~(\ref{eq12}) and (\ref{eq13}) leads to:
\[\chi '(\omega )=-\frac{c}{\omega_p}\alpha _o \exp\left({-\zeta\frac{\omega ^2}{\Delta _{in}^2}}\right)\mathrm{Erfi}\left({-\sqrt{\zeta}\frac{\omega }{\Delta _{in}}}\right)\] and:
\[\chi "(\omega )=-\frac{c}{\omega_p }\alpha _o\exp \left({-\zeta\frac{\omega ^2}{\Delta _{in}^2}}\right)\]
where $\alpha_o=\frac{2\pi b}{\delta }\frac{\beta g}{\Delta_{in}}$. In the central region of the absorption profile, dispersion $\omega_p \chi
'(\omega )/c$ is close to the linear approximation $\rho\alpha _o\omega /\Delta_{in} $, with $\rho \cong 1.385$, and $\vert \omega_p \chi '(\omega
)/c\vert$ is equal to the absorption coefficient at $\vert \omega/\Delta _{in} \vert = 0.425$.

\section{Signal recovery}\label{sec:recovery}
As already pointed out, the macroscopic coherence revives periodically, giving rise to delayed responses. In the framework of filtering by an
infinitely broad AFC, an exact replica of the incoming signal can be retrieved in backward direction at time $2\pi/\delta$ with 100\%
efficiency~\cite{afzelius2009}. This way, AFC storage appears to be perfectly reversible. Emission in backward direction involves two
counterpropagating $\pi$-pulses. They convert the optical atomic coherence into a long-lived coherence and back, extending the storage time well
beyond $2\pi/\delta$. When restored by the second $\pi$-pulse, the optical coherence phase has been changed by $2kz$, which causes emission in
backward direction. In the case of CRIB, perfect reversibility in backward direction can be achieved without the restriction of infinitely broad
processing bandwidth~\cite{moiseev2004}. We precisely aim at clarifying this point in the case of AFC.

Following the lines of refs~\cite{afzelius2009,moiseev2004}, we restrict the discussion to the backward retrieval scheme. Let
$\mathbf{A}_r(\tilde{\tau},z)$ represent the retrieved field in terms of the new moving frame coordinates $\tilde {\tau }=t+z/c$, $z=z$. In this frame
the field equation reads as:
\begin{equation}
\label{eq16}
-\textstyle{\frac{\partial}{\partial z}}\mathbf {A}_r (\tilde {\tau },z)=i\beta\exp \{i\omega _o \tilde {\tau }\}\mathbf {S}(\tilde {\tau },z),
\end{equation}
where $\mathbf {S}(\tilde {\tau },z)$ is comprised of the first revival of the atomic coherence previously excited by the signal field, and of the
instantaneous response of the medium to the retrieved field. Components $\mathbf{S}^{(1)}(\tilde {\tau},z)$ and $\mathbf{S}^{(0)}(\tilde {\tau },z)$, as defined in Eq.(\ref{eq9a}), respectively describe those two contributions. Substitution in Eq.(\ref{eq16}) leads to:
\begin{align}
\label{eq17}
-\textstyle{\frac{\partial} {\partial z}}\tilde {\mathbf {A}}_r (\omega,z)=&\frac{\omega_p }{2c}\{i\chi (\omega )\tilde {\mathbf {A}}_r (\omega,z)+
\nonumber \\ &
2\tilde {\kappa }(2\pi /\delta)\chi "(\omega )\}\tilde {\mathbf {A}}_p(\omega ,z),
\end{align}
\noindent
Solving in $z$ and Fourier transforming back to the time domain leads to the retrieved field expression at the medium output, i.e. in the front side at $z=0$:
\begin{equation}
\label{eq:retrieved}
\mathbf {A}_r(\tilde {\tau },0)=\tilde {\kappa }(2\pi /\delta)\int_{-\infty }^\infty {\frac{d\omega }{2\pi }}\Gamma(\omega) \tilde {\mathbf {A}}_p(\omega ,0)\exp(-i\omega \tilde {\tau }),
\end{equation}
where the complex efficiency $\Gamma(\omega )$ reads as:
\begin{equation}
\label{eq:efficiency}
\Gamma(\omega )=\frac{ 1-\exp [i\omega_p \chi (\omega )L/c]}{\{1-i\chi '(\omega )/\chi "(\omega )\}}.
\end{equation}
\noindent
With absorbing pikes much narrower than their spacing, $\tilde {\kappa }(2\pi /\delta)$ is close to $1$. Further assuming $G(\Delta)$ to be continuous , one can make factor $\{1-\exp [i\frac{\omega }{c}\chi(\omega )L]\}$ arbitrarily close to $1$ over the signal spectral range by increasing the medium
thickness $L$. Still the signal is not recovered with 100{\%} efficiency.

The frequency dependent weight factor $1/\left[1-i\chi '(\omega )/\chi"(\omega )\right]$ distorts and depletes the retrieved signal. The quantity $\chi'(\omega)/\chi"(\omega)$ represents nothing but the spatial phase mismatching that builds up over a round trip in the medium, down to the inverse absorption coefficient corresponding to the penetration depth. All the spectral components are depleted independently. Indeed, the phase mismatch builds up between the atoms and the signal field, at each frequency.

At this stage, the difference with CRIB can be clearly understood. In AFC-based memory, the part of information in channel $\omega$ that is captured at depth $z$ undergoes the frequency dependent $\omega_p\chi'(\omega)z/c$ phase shift, down to the storage position. The same phase shift is accumulated on the way back to the input surface when the signal is restored, which doubles the phase shift. In CRIB, the same superluminal feature takes place. However, information captured in channel $\omega$ is transferred to channel $-\omega$ when the frequency detuning is reversed. Hence, because $\chi'(\omega)$ is an odd function of $\omega$, the information initially captured at $\omega$, and retrieved at $-\omega$, undergoes phase shift $-\omega_p\chi'(\omega)z/c$ on the way back to the material surface, which cancels the phase shift accumulated on the way in~\cite{moiseev2004}. This is made possible because the detuning reversal does not affect propagation, at least when the inhomogeneously broadened atomic spectral distribution is symmetric with respect to the reversal center. Hence the susceptibility remains unchanged all along the process.

Despite the non-reversibility of AFC-based memory, the phase shift vanishes in the infinitely broad AFC limit considered in previous works. However, retrieval may be significantly altered when the signal bandwidth gets similar to the AFC width. Indeed $\chi'(\omega)/\chi"(\omega)$ variation over the AFC spectrum may exceed unity. For instance, in the frame of previously considered AFC gaussian envelope, $\chi'(\omega)/\chi"(\omega)\cong\rho\omega/\Delta_{in}$ for $|\omega| /\Delta _{in}<1$. This may strongly affect both the quantum efficiency and fidelity, as discussed in the following.

\subsection{Quantum efficiency}
The memory quantum efficiency (QE) is defined as:
\begin{equation}
Q=\frac{\int_{-\infty}^\infty d\omega\left\langle\tilde{\mathbf{A}}_r^+(\omega,0)\tilde{\mathbf{A}}_r(\omega,0)\right\rangle }
{\int_{-\infty}^\infty d\omega\left\langle\tilde{\mathbf{A}}_p^+(\omega,0)\tilde{\mathbf{A}}_p(\omega,0)\right\rangle }
\label{eq:MQE}
\end{equation}
where quantum averaging is performed over the initial state of light field. The spectral density of the input light intensity can be expressed as
$I_p(\omega)=\left\langle\tilde{\mathbf{A}}_p^+(\omega,0)\tilde{\mathbf{A}}_p(\omega,0)\right\rangle$. Then, substitution of Eqs (\ref{eq:retrieved})
and (\ref{eq:SPS}) into Eq. (\ref{eq:MQE}) leads to:
\begin{equation}
\label{eq20}
Q=[\tilde {\kappa }(2\pi /\delta )]^2\tilde {Q},
\end{equation}
where:
\begin{equation}
\label{eq:QE}
\tilde {Q}={\int_{-\infty} ^\infty {d\omega }\vert\Gamma (\omega )\vert^2I_p (\omega)} \mathord{\left/ {\vphantom{{\int_{-\infty }^\infty
{d\omega } \frac{I_p (\omega )}{\{1+[\chi'(\omega )/\chi "(\omega )]^2\}}} {\int_{-\infty }^\infty {d\omega }
I_p (\omega )}}} \right. \kern-\nulldelimiterspace} {\int_{-\infty}^\infty {d\omega } I_p (\omega )},
\end{equation}
\noindent
Quantum efficiency critically depends on the narrowness of the absorbing pikes through factor $[\tilde {\kappa }(2\pi /\delta) ]^2$ that approaches 1 when $b<<\delta$.

As an illustration, let a Gaussian shaped signal $I_p (\omega )=I_p (0 )\sqrt{\frac{\zeta}{\pi}}\frac{1}{\Delta_p} \exp \left({-\zeta\frac{\omega ^2}{\Delta _p^2}}\right)$ be captured through the previously considered Gaussian shaped AFC. Under assumption of infinite optical depth, QE can be expressed for $\Delta_p/\Delta_{in}<1$ as:

\begin{align}
\label{eq22}
Q_G =&\frac{[\tilde {\kappa }(2\pi /\delta) ]^2\sqrt {\pi /2} }{0.9\vert \Delta_p /\Delta _{in} \vert }\exp \{\frac{1}{2(0.9\Delta_p /\Delta _{in})^2}\}\nonumber \\&\mathrm{Erfc}\{\frac{1}{0.9\sqrt 2 \vert \Delta_p /\Delta _{in} \vert }\}.
\end{align}
\noindent
\begin{figure}
  \includegraphics[width=0.4\textwidth]{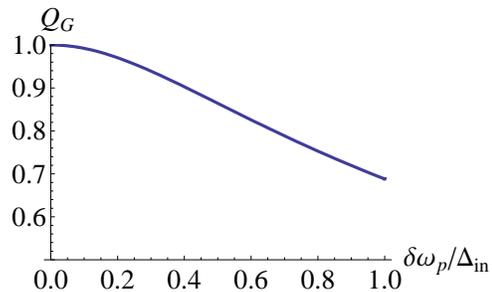}
  \caption{Quantum efficiency $Q_G $ as a function of relative bandwidth $\omega _p /\Delta _{in} $ when both the AFC and the captured signal are Gaussian shaped. Atomic decoherence is ignored and the optical depth is infinite.}
  \label{Figure2}
\end{figure}
The corresponding QE is plotted in Fig.\ref{Figure2} as a function of the relative bandwidth $\Delta_p /\Delta _{in} $. It appears that QE drops below 90\% when the signal bandwidth exceeds 0.3 times the AFC width. The decrease can be even stronger for different spectral shapes, such as a Lorentzian.

The extension to multiple temporal mode storage is straightforward when the different temporal modes do not overlap. Then multimode QE coincides with the single mode efficiency, provided the comb shape does not change during the capture of the mode train.

\subsection{Fidelity }\label{sec:fidelity}
The memory fidelity is expressed in terms of the output state projection on the input state. Let the signal field be prepared in a single photon state:
\begin{equation}
\left|{\psi_{in}(t)}\right\rangle =\int_{-\infty}^\infty{d\omega f_p (\omega )\exp \{-i\omega t\}\mathbf {a}^+(\omega )\left| 0\right\rangle }
\label{eq:SPS}
\end{equation}
where the creation and annihilation operators $\mathbf {a}^+(\omega )$ and $\mathbf {a}(\omega ')$ obey the commutation relation $[\mathbf {a}(\omega
'),\mathbf {a}^+(\omega )]=\delta (\omega -\omega ')$, and where the spectral distribution $f_p (\omega )$ satisfies the normalization condition:
\[\int_{-\infty }^\infty{d\omega \vert f_p (\omega )\vert ^2=1}.\] The normalized output state can be derived from Eq. (\ref{eq:retrieved}) as:
\begin{align}
\label{eq:output_state}
\left|\psi_{out}(t)\right\rangle= &\int_{-\infty}^\infty d\omega \Gamma(\omega)f_p (\omega )\exp[-i\omega (t-T)] \mathbf {a}^+(\omega )\left| 0\right\rangle
\nonumber\\ & \left/{\left[\int_{-\infty}^\infty d\omega\vert\Gamma(\omega)f_p(\omega)\vert^2\right]^{1/2}}\right.
\end{align}
where $T$ represents the total storage duration. Then, defining fidelity as $F=\vert\left\langle\psi_{out}(t)\left|\psi_{in}(t-T)\right\rangle\vert^2 \right.$, one readily obtains:
\begin{align}
\label{eq:fidelity}
F= \left|\int_{-\infty}^\infty d\omega \Gamma(\omega)\vert f_p (\omega ) \vert^2\right|^2 \left/{\int_{-\infty}^\infty d\omega\vert\Gamma(\omega)f_p(\omega)\vert^2}\right.
\end{align}

Fidelity can be optimized independently of QE. Fidelity does not depend on pike narrowness, instead of QE (see Eq. (\ref{eq20})). Moreover, while QE drops to zero at small optical density, fidelity can approach 100\% in the same conditions. Indeed, $\Gamma(\omega)\cong\omega_p\chi"(\omega)L/c$ when $\omega_p\chi"(\omega)L/c<<1$ and, provided absorption is uniform over the signal bandwidth, substitution in Eq. (\ref{eq:fidelity}) leads to $F\cong100\%$, which is consistent with recent experimental results~\cite{clausen2011,sagla2011}.

In the extreme opposite conditions of infinite optical density, a situation of interest is offered by symmetric spectral shapes, with $\vert f_p(\omega)\vert^2=\vert f_p(-\omega)\vert^2$. Broadband symmetric shapes should indeed be used to make the best of the memory multimode capacity. Then, since $\chi'(\omega)=-\chi'(\omega)$, Eq. (\ref{eq:fidelity}) reduces to:
\begin{equation}
\label{eq:fidelity_high_OD}
F = \int\limits_{-\infty }^\infty {d\omega } \frac{\vert f_{p,s} (\omega)\vert ^2 }
{\{1+[\chi '(\omega )/\chi "(\omega )]^2\}}.
\end{equation}
\noindent
In this situation, phase mismatching affects fidelity in quite the same way as QE (see Eq. (\ref{eq:QE})).

In summary, phase mismatching significantly alters both quantum efficiency and fidelity of finite bandwidth AFC quantum memory. In the next section we propose a method to compensate dispersion effects in order to take better advantage of the AFC bandwidth and of the memory multimode capacity.

\section{Modified AFC scheme with compensated spectral dispersion }\label{sec:MAFC}
\begin{figure}
  \includegraphics[width=0.4\textwidth,height=0.2\textwidth]{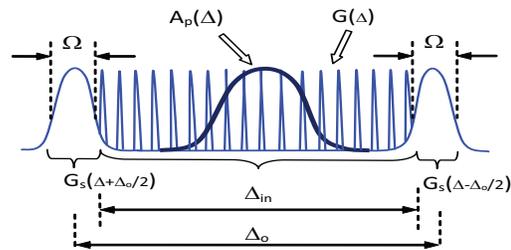}
  \caption{Modified AFC. The AFC is contained between two absorbing lines. Those lines give rise to positive dispersion and slow light effect that compensate the AFC-induced negative dispersion and fast light feature.}
  \label{Figure3}
\end{figure}
As discussed above, the AFC memory is affected by spectral dispersion. Dispersion leads to superluminal propagation features and to phase mismatching of the retrieved field. Phase mismatching reflects incomplete light-atoms reversibility, in contrast with the CRIB scheme~\cite{moiseev2004}. In order to restore reversibility, we propose a modified AFC (MAFC) scheme that cancels the dispersion effect.

While fast light features spoil AFC memory reversibility, the opposite slow light effect can also be observed in the context of linear absorption, as discussed~\cite{shakhmu2005} and experimentally demonstrated recently~\cite{camacho2006,lauro2009,walther2009}. Indeed slow light is not specific to non-linear processes such as electromagnetically induced transparency (EIT), but can also merely result from the existence of a transparency window within an absorption profile. This suggests us to compensate the AFC superluminal effect by simply inserting the memory comb within the transparent spectral interval between two aborbing lines.
\begin{figure}
  \includegraphics[width=0.45\textwidth]{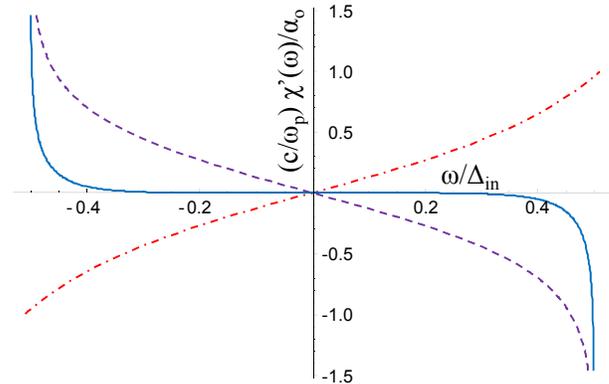}
  \caption{variations of $\chi _A'(\omega )$ (dashed line), $\chi _s'(\omega )$ (dashed-dotted line) and $\chi _M'(\omega )$ (solid line) over the AFC spectral range. The AFC fast light effect (negative dispersion) is compensated by the slow light effect (positive dispersion) induced by the side lines.}
  \label{fig:slow_fast}
\end{figure}

The MAFC principle is sketched in Fig. \ref{Figure3}. The AFC occupies the central part of the absorption profile. A rectangular shape is assigned to the $G_{o}(\Delta )$ AFC envelope. This function is flat over the $\Delta_{in}$-wide memory bandwidth and drops to zero outside this interval. On the sides of the AFC, the absorption lines $G_{s}(\Delta +\Delta _{o}$/2) and G$_{s}(\Delta -\Delta_{o}$/2) generate the slow light effect that will cancels the fast light feature induced by the AFC. The line centers are separated by interval $\Delta_{o}$. Spectral hole burning techniques~\cite{lauro2009,usmani2010,bonarota2010,clausen2011} can be used to shape such a complex absorption profile.

Let the normalized rectangular AFC envelope be defined as:
\begin{equation}
\label{eq30}
G_o (\Delta )=\left\{ {\begin{array}{l} \frac{1}{\Delta _{in} },\vert\Delta\vert <\Delta _{in} /2 \\
\  \   0\ ,\vert\Delta\vert >\Delta _{in} /2 \\
 \end{array}} \right.,
\end{equation}
The absorption lines on the sides are given the following normalized shape:
\begin{equation}
\label{eq31}
G_{s} (\Delta \pm \Delta _o /2)=\frac{\Omega }{\pi [(\Delta \pm \Delta _o /2)^2+\Omega^2]},
\end{equation}
\noindent
Rectangular shaped AFC was typical in recent experiments~\cite{deried2008,usmani2010,bonarota2010,clausen2011,sagla2011}. However the role of the side parts of the absorption profile was not discussed.

We aim at making the real part of susceptibility vanish over the AFC. We rely on two adjustable parameters, namely the relative amplitude $f_s$ of the side lines and their relative spacing $\Delta_o/\Delta_{in}$, to optimize the suppression of dispersion effects.
\begin{figure}
  \includegraphics[width=0.45\textwidth]{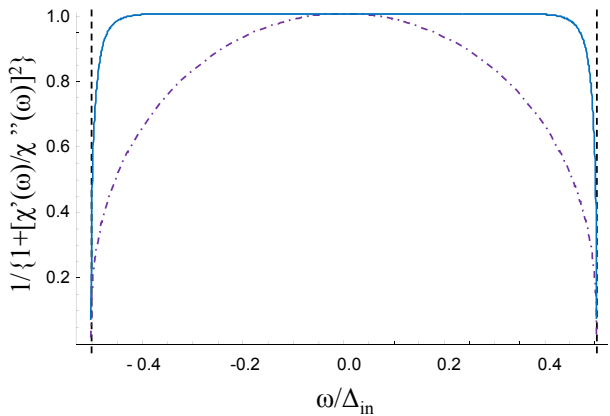}
  \caption{AFC (dashed-dotted line) and MAFC (solid line) spectral distribution of $1/\left(1+[\chi'(\omega )/\chi''(\omega )]^2\right)$ over the AFC bandwidth. }
  \label{fig:phase_mismatch}
\end{figure}

One should be aware that the present AFC departs from our assumptions in sections~\ref{sec:capturing} and \ref{sec:recovery}. In those sections the
AFC envelope was represented by a continuous function. Therefore one could make the optical density arbitrarily large over the entire processing
spectral range by increasing the medium thickness $L$. In the present case, the AFC is sharply cut on the edges. Therefore signal wings outside the
$\Delta_{in}$-wide band are unavoidably lost. Another consequence of the AFC rectangular shape is to invalidate the slowly varying envelope assumption
we made in section~\ref{sec:capturing}, and especially in Eq. (\ref{eq7}). However this only affects the very edges of the AFC, over a spectral range
of order $b<<\Delta_{in}$. Hence we shall neglect the corresponding error.

Let us examine the signal transmission over the AFC spectral interval $\left[-\Delta_{in}/2,\Delta_{in}/2\right]$. According to Eqs. (\ref{eq12}-\ref{eq:alpha}), the real and imaginary parts of AFC susceptibility read as:
\begin{equation}
\chi _A '(\omega )=\frac{c}{\omega_p}\alpha_o\frac{1}{\pi}\mathrm{Ln}\left(\frac{\Delta _{in} /2-\omega }{\Delta _{in} /2+\omega}\right)
\end{equation}
\begin{equation}
\chi _A ''(\omega )=-\frac{c}{\omega_p}\alpha_o,
\end{equation}
where
\begin{equation}
\label{eq:alpha_o}
\alpha_o=2\pi\frac{\beta g}{\Delta_{in}}\frac{1}{\kappa (0)\delta }.
\end{equation}
Under assumption $\Omega<<\Delta_o-\Delta_{in}$, the side lines stay outside the AFC. Residual absorption by those lines at the AFC center is expressed by the absorption coefficient:
\begin{equation}
\label{eq:alpha_s}
\alpha_s(0)=4f_s\frac{\beta g\Omega}{\left(\Delta_o/2\right)^2}.
\end{equation}
This does not affect the signal round trip through the AFC medium provided:
\begin{equation}
\label{eq:alpha_s_ignored}
\frac{2\alpha_s(0)}{\alpha_o}=\frac{4}{\pi}f_s\kappa (0)\delta\frac{\Omega\Delta_{in}}{\left(\Delta_o/2\right)^2}<<1.
\end{equation}

As for the real part of susceptibility, the side lines bring the contribution:
\begin{equation}
\chi _s '(\omega )=\frac{c}{\omega_p}\alpha_o\frac{1}{\pi}f_s\kappa(0)\delta  \frac{2\omega \Delta _{in} }{(\Delta _o /2)^2-\omega^2}
\end{equation}
The real part of modified AFC susceptibility reads as $\chi _M '(\omega )=\chi _A '(\omega )+\chi _s '(\omega )$ and vanishes to third order in $\omega/\Delta_{in}$ provided:
\begin{align}
\label{eq:compensation}
\Delta_o=\sqrt{3}\Delta_{in},\  f_s\kappa(0)\delta=\frac{3}{2 }
\end{align}

The spectral variations of $\chi _A'(\omega )$, $\chi _s'(\omega )$ and $\chi _M'(\omega )$ are displayed in Fig. \ref{fig:slow_fast}, showing the compensation of AFC fast light by the slow light effect from the side-lines. As a result, in the $\alpha_oL>>1$ large optical density limit, the efficiency factor $\vert\Gamma(\omega)\vert^2$ of the modified AFC gets close to unity over the entire AFC bandwidth, unlike the original AFC scheme, as illustrated in Fig. \ref{fig:phase_mismatch}.

\begin{figure}
  \includegraphics[width=0.45\textwidth]{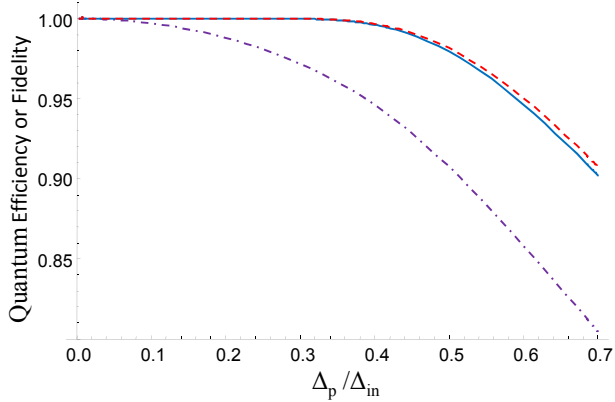}
  \caption{Gaussian signal recovery. Quantum efficiency and/or fidelity are displayed as a function of the signal spectrum FWHM, normalized to the AFC bandwidth. AFC (dashed-dotted line) and MAFC (solid line) expected accomplishments are displayed together with perfect operation result (dotted line), with complete elimination of dispersion. }
  \label{fig:QE_or_F}
\end{figure}

Dispersion cancellation impacts on quantum efficiency and fidelity, as illustrated in Fig.~\ref{fig:QE_or_F}. In this figure, the incoming signal is assigned a gaussian spectrum, centered at the middle of the AFC band, and the optical density is assumed to be large. As discussed in Section~\ref{sec:fidelity}, quantum efficiency and fidelity are expressed by the same number in this situation. Unlike conventional AFC parameters, MAFC quantum efficiency and fidelity keep very close to an ideal process, with complete elimination of dispersion. However, as pointed out above, both quantum efficiency and fidelity suffer from the loss of spectral components lying outside the AFC processing band, even when dispersion effects completely cancel.

\section{Discussion and conclusion}
Infinite bandwidth, AFC-based, quantum memory may offer 100\% quantum efficiency and fidelity when the signal is retrieved in backward direction. This is no longer true when the memory spectral width is limited. Since the dispersive part of susceptibility no longer vanishes, the retrieved signal is no longer spatially phase-matched to the atomic coherences. In contrast with CRIB, the mismatching accumulated during the storage step is not compensated during the recovery stage, which reveals the intrinsic non-reversibility of AFC.

After demonstrating these features, we have proposed a modified AFC scheme that restores spatial phase matching and reversibility. By setting absorption lines on the sides of the AFC, we reduce the group velocity. Hence, the superluminal effect caused by the AFC itself is cancelled by a slow light effect.

The analysis, and the method to cure non-reversibility, could be extended to recently proposed photon echo~\cite{moiseev2011,mcauslan2011,damon2011} or Raman echo~\cite{hetet2008b,nunn2008,legouet2009,hosseini2009,moiseev2011b} QM schemes based on persistent inhomogeneous broadening.

\section{Acknowledgments}

SAM would like to thank Universit\'e Paris Sud for helping to support his visit to Laboratoire Aim\'{e} Cotton and for the hospitality shown to him during his visit. SAM also acknowledges the support from Russian Foundation for Basic Research through grant {\#} 10-02-01348.

\bibliography{MAFC}

\end{document}